\documentclass[pra,twocolumn,showpacs,floatfix,amsfonts,nofootinbib]{revtex4}
\usepackage{bm}
\usepackage{graphicx}
\usepackage{here}
\usepackage{latexsym}
\usepackage{amsmath}
\usepackage{amssymb}
\usepackage{dcolumn}
\usepackage{ifthen}
\usepackage{color}

\begin{document}

\newcommand{\x}{{\bf x}}
\renewcommand{\k}{{\bf k}}
\newcommand{\E}{{\bf E}}
\newcommand{\e}{{\bf e}}
\def\mn#1{\marginpar{*\footnotesize \footnotesize #1}{*}}
\def\isum{{\sum\!\!\!\!\!\!\!\!\int}}
\def\bfzero{{\bf 0}}
\newcommand{\ket}[1]{\, | #1 \rangle}
\newcommand{\be}{\begin{equation}}
\newcommand{\ee}{\end{equation}}
\newcommand{\bea}{\begin{eqnarray}}
\newcommand{\eea}{\end{eqnarray}}
\newcommand{\bdm}{\begin{displaymath}}
\newcommand{\edm}{\end{displaymath}}
\newcommand{\bfd}{{\bf d}}
\newcommand{\bfk}{{\bf k}}
\def\hF{{\bar F}}
\newcommand{\hbE}{{\hat {\bf E}}}
\newcommand{\hbP}{{\hat {\bf P}}}
\newcommand{\hpsi}{{\hat {\psi}}}
\renewcommand{\k}{{\bf k}}
\renewcommand{\r}{{\bf r}}
\newcommand{\bfp}{{\bf p}}
\newcommand{\bfq}{{\bf q}}
\newcommand{\bfE}{{\bf E}}
\newcommand{\bfP}{{\bf P}}
\newcommand{\ha}{{\hat a}}
\newcommand{\bfr}{{\bf x}}
\newcommand{\bfz}{{\bf 0}}
\newcommand{\cl}{{\cal l}}
\newcommand{\cE}{{\cal E}}
\newcommand{\tG}{{\widetilde G}}
\newcommand{\veps}{\varepsilon}
\newcommand{\non}{\nonumber \\}
\newcommand{\mGamma}{\bar \Gamma}

\def\gsym{\gamma}

\def\pder#1#2{\frac{\partial #1}{\partial #2}}
\def\lsim{\raise0.3ex\hbox{$<$\kern-0.75em\raise-1.1ex\hbox{$\sim$}}}
\def\Type{P}
\def\LorP#1#2{
\ifthenelse{\equal{\Type}{L}}{#1}{#2}
}

\title{Sequential superradiant scattering from atomic Bose-Einstein condensates}
\author{O.\ Zobay and G.~M.\ Nikolopoulos}
\affiliation{
\LorP{\vspace*{0.2cm}}{}
Institut f\"ur Angewandte Physik, Technische Universit\"at Darmstadt, 
\LorP{\\ Hochschulstr.\ 4a,}{\hspace*{-0.4cm}}
64289 Darmstadt, Germany
\LorP{\\ E-Mail: oliver.zobay@physik.tu-darmstadt.de \vspace*{0.2cm} }{}
}
\date{\today}
\begin{abstract}
We theoretically discuss several aspects of sequential superradiant scattering from atomic Bose-Einstein condensates. Our treatment is based on the semiclassical description of the process in terms of the Maxwell-Schr\"odinger equations for the coupled matter-wave and optical fields. First, we investigate sequential scattering in the weak-pulse regime and work out the essential mechanisms responsible for bringing about the characteristic fan-shaped side-mode distribution patterns. Second, we discuss the transition between the Kapitza-Dirac and Bragg regimes of sequential scattering in the strong-pulse regime. Finally, we consider the situation where superradiance is initiated by coherently populating an atomic side mode through Bragg diffraction, as in studies of matter-wave amplification, and describe the effect on the sequential scattering process.
\end{abstract}
\maketitle

\LorP{ \hspace*{0.3cm} {\bf Topic: Physics of Cold Trapped Atoms\\[0.3cm]
\hspace*{0.3cm} Proceedings of LPHYS'06 workshop:\\
\hspace*{0.3cm} Paper 6.8.4}
\newpage \textcolor{white}{.} \newpage }{}

\section{Introduction}
\label{sec:Int}

The recent observation of superradiant scattering from atomic Bose-Einstein condensates (BECs) \cite{InoChiSta99,SchTorBoy03} has opened up a new perspective on collective processes involving matter and light. In a typical superradiance experiment, an elongated BEC is exposed to a far detuned laser pulse with wave vector $k_l{\bf e}_x$ traveling perpendicular to the condensate axis. This ``pump pulse'' causes some of the atoms to undergo spontaneous Rayleigh scattering and acquire recoil momentum. Due to interference, the recoiling atoms and the BEC atoms at rest start to form matter-wave gratings whose amplitudes grow rapidly in a self-amplifying process induced by subsequent scattering events. Initially, the scattered photons can go in any direction; however, due to phase matching effects induced by the elongated condensate shape, the fastest growth is experienced by those gratings for which the scattered photons travel up and down the long condensate axis with wave vectors $\pm k{\bf e}_z$ in the two so-called optical endfire modes. Following a mode-competition process \cite{MooMey99}, the ensuing superradiant light emission is concentrated into the two endfire modes. Consequently, the recoiling atoms possess very well-defined momenta $k_l{\bf e}_x\pm k{\bf e}_z$ making 45$^{\rm o}$ angles with the direction of the applied laser pulse, and they thus form two first-order atomic ``side modes." To observe the side modes, one lets the system evolve freely for some time after the laser pulse has been turned off. During this time the recoiling atoms separate in space from the condensate at rest and become clearly visible as distinct spots in absorption images. 

The above scenario describes the basic mechanism of BEC superradiance, but the full range of effects turns out to be much richer. On the one hand, BEC atoms can also absorb a photon from an endfire mode and re-emit it into the laser mode. The accompanying momentum transfer $-k_l{\bf e}_x\pm k{\bf e}_z$ causes the atoms to recoil backwards at a 45$^{\rm o}$ angle against the direction of the applied laser field. However, this process violates energy conservation and only occurs for intense laser pulses shorter than the recoil time (Kapitza-Dirac regime). On the other hand, upon an increase in the pump-pulse duration the atoms in the side modes may undergo additional scattering cycles thereby populating second- and higher-order side modes (sequential scattering). In the Kapitza-Dirac regime, one typically observes an X-shaped pattern with the initial BEC in the center and the recoiling atoms moving both into and against the direction of the applied laser pulse \cite{SchTorBoy03}. For weaker pulses with a duration long compared to the recoil time (Bragg regime), however, the side modes are observed to form a characteristic fan pattern involving forward recoiling atoms only \cite{InoChiSta99}.

In recent work, we have derived a theoretical description that is able to reproduce and explain these and other key observations of BEC superradiance in a unified way \cite{ZobNik05,ZobNik06}. Our approach which is based on the time-dependent Maxwell-Schr\"odinger equations follows two main concepts. First, we describe the field dynamics semiclassically. This treatment can be expected to be very accurate as soon as the populations of side and endfire modes become macroscopic and quantum fluctuations are no longer essential. The second main feature is the inclusion of spatial propagation effects of the matter-wave and optical fields along the condensate axis. It turns out that the consideration of this aspect is crucial to explain some of the main observations of the experiments \cite{InoChiSta99,SchTorBoy03} and to obtain a complete understanding of BEC superradiance. While spatial effects have also been considered in Refs.\ \cite{AveTri04,UysMey06}, the majority of the theoretical literature on BEC superradiance is based on spatially-independent models where propagation effects are not taken into account (see, e.g., Refs.\ \cite{MooMey99,MusYou00,BenBen04,VasEfiTri04,RobPioBon04}). Nevertheless, the spatially independent description is still useful to investigate some basic aspects of the physics underlying superradiant scattering from BECs, and in this paper we will apply both types of models in our analyses.

The purpose of this paper is to discuss some topics relevant to sequential superradiance that were not covered in our previous work \cite{ZobNik05,ZobNik06}. After briefly reviewing the spatially dependent and independent semiclassical descriptions of BEC superradiance in Sec.\ II, we turn to the study of sequential scattering in the Bragg regime in Sec.\ III. It has already been established in Refs.\ \cite{ZobNik05,ZobNik06} that the characteristic fan-shaped side-mode distribution patterns observed in this regime \cite{InoChiSta99} can be reproduced with the help of the semiclassical models. In this paper we would like to discuss several simple, yet important physical mechanisms that we think essential for a conceptual understanding of the system behavior in the Bragg regime. In particular, they will be useful in explaining the emergence of the characteristic fan patterns. 
More specifically, the principles that we will discuss are the following: first, the concept of ``self-adapting" Bragg scattering, second, the role of the detuning barriers related to the transitions between different atomic side modes, and third, a directionality in the population transfer between side modes that favors the transition along the direction of the pump pulse rather then against it.
Self-adapting Bragg scattering refers to the fact that the optical endfire modes do not contain only a single frequency resonant with the transition between the BEC at rest and the first-order forward side modes. Rather, each transition between side modes contributes its own resonant frequency component to the endfire mode fields. The creation of these components ensures the continuation of the superradiant scattering process. 

A very interesting question regarding the nature of the Kapitza-Dirac regime in superradiant scattering was raised in Ref.\ \cite{RobPioBon04}. Whereas Ref.\ \cite{SchTorBoy03} claimed that Kapitza-Dirac scattering is essentially connected to the duration of the pump pulse and only occurs at short enough times, Ref.\ \cite{RobPioBon04} stated that the pulse strength should be the distinguishing feature. We would like to contribute to this discussion by numerically examining the regime of strong pulses of longer duration. Our studies that are discussed in Sec.\ IV indicate that for sufficiently long times, although the system shows a complex behavior, one can indeed observe an overall tendency for a transition towards the Bragg scattering regime.
Apart from spontaneous Rayleigh scattering, BEC superradiance can also be initiated by transferring a small amount of population into one of the side modes by means of Bragg diffraction. This seeding mechanism has been used in studies of matter-wave amplification \cite{InoPfaGup99,SchCamStr04,KozSuzTor99}.
In Sec.\ V, we investigate how the seeding of a side mode modifies the sequential superradiant scattering. 
The paper concludes with a short summary in Sec.\ VI.

\section{Semiclassical theory of superradiant scattering}

In this section we briefly summarize our theory of superradiant Rayleigh scattering developed in Refs.\ \cite{ZobNik05,ZobNik06}. As outlined in the Introduction, we consider an elongated condensate of length $L$ and oriented along the $z$ axis that is exposed to a far-detuned pump laser pulse $\cE_l(t){\bf e}_y (e^{i(k_lx-\omega_l t)}+c.c.)/2$ with $\omega_l = ck_l$, traveling in the $x$ direction. Collective effects in the interaction between the atoms and the pulse lead to superradiant light scattering into the optical endfire modes $\cE_\pm(z,t)\e_y (e^{-i(\omega t\mp kz)}+c.c.)$. As a consequence, the recoiling atoms have well-defined momenta and appear in distinct atomic side modes. In the side mode indexed $(n,m)$, 
atoms possess momentum $\hbar(nk_l\e_x + mk\e_z)$, while their kinetic 
energy is given by $\hbar\omega_{n,m} \approx \hbar(n^2+m^2)\omega_r$ with $\omega_r = \hbar k_l^2/2M$ the atomic recoil frequency. Since the pump pulse is far-off resonant, we can adiabatically eliminate the excited atomic state and only need to consider atoms in their internal ground state.

To derive equations of motion for the optical and matter-wave fields $\E(\x,t)$ and $\psi(\x,t)$, we decompose them according to
\bea\label{svea_psi}
\psi(\x,t) &=& \sum_{(n,m)} \frac{\psi_{nm}(z,t)}{\sqrt A} e^{-i(\omega_{n,m}t-nk_lx - mkz)},\\ \label{svea_ep}
\E^{(+)}(\x,t) &=& \cE_l\e_y e^{-i(\omega_l t-k_lx)}/2 +\cE_+(z,t)\e_y e^{-i(\omega t-kz)}\non
&& + \cE_-(z,t)\e_y e^{-i(\omega t+kz)},
\eea
with $\E^{(+)}$ denoting the positive-frequency component of the electric field, and $\psi_{nm}(z,t)$ and $\cE_\pm(z,t)$ the slowly-varying field envelopes along the $z$ axis. The average condensate cross section is denoted $A$, and $\omega=kc$. The sum of Eq.\ (\ref{svea_psi}) contains only terms with even $n+m$ since we only consider atoms in their internal ground state. In order to introduce a concise notation, we rescale the optical and matter waves 
as  
\bea
\cE_{\pm,l} &\to&  e_{\pm,l}\sqrt{\frac{\hbar\omega k_l}{2\veps_0 A}};\quad 
\psi_{nm} \to \frac{\psi_{nm}\sqrt{k_l}}{\sqrt{A}}
\eea
and define the dimensionless time $\tau=2\omega_{r}t$ and length $\xi=k_lz$.
Starting from the full Maxwell-Schr\"odinger equations for the coupled electric and matter-wave fields \cite{CohDupGry89,ZhaWal94,Mey01}, we can use the expansions (\ref{svea_psi}) and (\ref{svea_ep}) together with the slowly-varying envelope approximation to derive the equations of motion for the envelope functions $\psi_{nm}$ and $\cE_\pm$. In this way, we find \cite{ZobNik05,ZobNik06}
\begin{subequations}
\label{sdeq}
\bea\label{env_psi}
&&i\pder{\psi_{nm}(\xi,\tau)} \tau = -\frac 1{2} \pder{^2 \psi_{nm}}{\xi^2} 
-i m \pder{\psi_{nm}}\xi\\
&&+\kappa\big[e_+^*\psi_{n-1,m+1}e^{i(n-m-2)\tau} 
+e_-^*\psi_{n-1,m-1} e^{i(n+m-2)\tau}\nonumber\\ 
&&+e_+\psi_{n+1,m-1}e^{-i(n-m)\tau}+e_-\psi_{n+1,m+1} e^{-i(n+m)\tau}\big] \nonumber
\eea
where the coupling constant between matter-wave and electric fields is given by
\bdm
\kappa = \frac{g}{2\omega_r}\sqrt{k_lL}
\edm
with
\bdm
g = \frac{|\bfd|^2\cE_l}{2\hbar^2\delta} 
\sqrt{\frac{\hbar\omega_l}{2\veps_0 AL}}.
\edm
Here, $\bfd$ denotes the atomic dipole moment between ground and excited state, $\delta$ the laser detuning from resonance, and $\veps_0$ the vacuum permittivity. Due to the small size of the condensate, we can neglect retardation effects in the propagation of the electric fields. Therefore, we can directly express the endfire mode envelopes in terms of the atomic side modes, i.e.,
\bea\label{e_p}
e_+(\xi,\tau) &=& -i\frac{\kappa}\chi\int_{-\infty}^\xi d\xi' 
\sum_{(n,m)} e^{i(n -m)\tau}\non
&& \times \psi_{nm}(\xi',\tau)\psi_{n+1,m-1}^*(\xi',\tau),\\
\label{e_m}
e_-(\xi,\tau) &=& -i\frac{\kappa}\chi\int^{\infty}_\xi d\xi' 
\sum_{(n,m)} e^{i(n+m)\tau}\non
&& \times \psi_{nm}(\xi',\tau)\psi_{n+1,m+1}^*(\xi',\tau)
\eea
\end{subequations}
with $\chi = \frac{ck_l}{2\omega_r}$.
From these equations one may clearly see how the build-up of the endfire-mode 
fields at $(\xi,t)$ is driven by the coherences $\psi_{nm}\psi_{n+1,m\pm 1}^*$.

The physical contents of Eq.\ (\ref{env_psi}) can be interpreted as follows. The first term on the right-hand side describes the 
quantum-mechanical dispersion of the envelope function, while the second one 
leads to a spatial translation with velocity $v_m = m\hbar k/M$. The remaining terms describe photon exchange 
between one of the endfire modes and the laser beam.
In particular, through 
stimulated scattering, an atom in a side mode $(n,m)$ can absorb a laser 
photon and deposit it into one of the endfire modes. 
The accompanying recoil transfers the atom into one of the side modes 
$(n+1,m\pm 1)$. Alternatively, the atom may absorb an endfire-mode photon and 
emit it into the laser beam, thereby ending up in the side mode 
$(n-1,m\pm 1)$. This latter process is causes atoms to recoil backward \cite{SchTorBoy03}.

In general, superradiant scattering is initiated by quantum-mechanical noise, i.e., spontaneous Rayleigh scattering from individual condensate atoms \cite{MooMey99}. Adapting the discussion of Refs.\ \cite{GroHar82,HaaKinSch79a,HaaKinSch79b} regarding ``conventional" superradiance, we can take the effects of the initial quantum-mechanical fluctuations into account by solving the semiclassical equations of motion with stochastic initial conditions (seeds) for the side modes $\psi_{1,\pm 1}(\xi,\tau=0)$. Since the noise is practically relevant for the initial population of those modes only, we can set $\psi_{nm}(\xi,\tau=0) = 0$ for all other side modes. The condensate $\psi_{00}$ is chosen to be in the macroscopic ground state at $t=0$.
In our simulations, we follow the experimental data of
Refs.\ \cite{InoChiSta99,SchTorBoy03} and consider a $^{87}$Rb BEC of $N=2\times 10^6$ atoms with length $L=200\mu$m and cross-section diameter $15\mu$m. 
The condensate is in the Thomas-Fermi regime, so that we can model its wave 
function as $\psi_{00}(z) = \sqrt{n(z)}$ with 
$n(z) = C[(L/2)^2-z^2]\Theta(L/2-|z|)$, $C=3N/4L^3$.
As in Refs.\ \cite{ZobNik05,ZobNik06}, the 
first-order atomic side modes are seeded according to 
$\psi_{1,\pm 1}(z,0) = \psi_{00}(z)/\sqrt N$, which corresponds to one 
delocalized atom in each of the modes. Our results are not qualitatively 
influenced by the choice for the shape of the seed function.
The applied laser pulse is modeled as rectangular 
lasting from $t=0$ up to $t=t_f$.

As mentioned in the Introduction, while the spatially dependent description derived above is essential for a complete understanding of superradiant scattering, basic aspects can already be studied within a simpler spatially independent model. The equations of motion for this model can formally be derived from Eqs.\ (\ref{sdeq}) by dropping any spatial dependence and setting $\psi_{nm}\to C_{nm}/\sqrt{k_l L}$ 
and $e_\pm\to b_\pm/\sqrt{k_l L}$. The amplitudes $C_{nm}(\tau)$ and $b_\pm(\tau)$ can be interpreted as amplitudes of the matter-wave and optical fields, respectively, and 
they obey the equations of motion
\begin{subequations}
\label{sieq}
\bea
\label{env_psi_si}
&&\!\!\!\!\!\!\frac{dC_{nm}(\tau)} {d\tau} = -i\frac{g}{2\omega_{\rm r}}\Big [b_+^*C_{n-1,m+1}e^{i(n-m-2)\tau}\non
&&  +b_-^*C_{n-1,m-1} e^{i(n+m-2)\tau}+b_+C_{n+1,m-1}e^{-i(n-m)\tau}\non
&&+b_-C_{n+1,m+1} e^{-i(n+m)\tau}\Big ],\\
\label{e_p_si}
&&\!\!\!\!\!\! b_+(\tau) = -i\frac{g}{\gamma}
\sum_{(n,m)} e^{i(n -m)\tau}C_{nm}(\tau)C_{n+1,m-1}^*(\tau),\\
\label{e_m_si}
&&\!\!\!\!\!\! b_-(\tau) = -i\frac{g}{\gamma}
\sum_{(n,m)} e^{i(n+m)\tau}C_{nm}(\tau)C_{n+1,m+1}^*(\tau)
\eea
\end{subequations}
with $\gamma = 2c/L$.

\section{Sequential scattering in the Bragg regime}

The Kapitza-Dirac and the Bragg regimes of superradiant scattering are characterized by the emergence of disctinctive X-shaped and fan-like side-mode distribution patterns, respectively. 
While the appearance of the X patterns can concisely be explained in terms of the spatial localization of the endfire and side modes at the BEC edges \cite{ZobNik05,ZobNik06}, the fan patterns show a significantly more complex behavior which appears to be determined by an interplay of different effects. In this section we would like to discuss three concepts which we think to be important in understanding the physics underlying sequential scattering in the Bragg regime.

{\it (i) Self-adapting Bragg scattering.} Following the onset of the superradiant process, the optical endfire modes contain only one frequency component $\omega=\omega_l-2\omega_r$ (here we are considering times larger than the recoil time so that frequencies can be well defined on the scale of $\omega_r$). However, as more and more side modes become occupied, we see from Eqs.\ (\ref{e_p}) and (\ref{e_m}) that additional frequency components are created. More specifically, we find that the simultaneous presence of population in the side modes $(n,m)$ and $(n+1,m\pm 1)$ produces a frequency component at $\omega_l-2(n\pm m+1)\omega_r$. This frequency is resonant with the transition between the two side modes.
 
However, to initiate the population transfer to an empty higher-order side mode, it is not necessary that resonant light is present in the endfire modes. Rather, small amounts of population can also be transferred in an off-resonant Bragg scattering process involving the pump laser, an occupied side mode, and detuned light in the endfire mode\footnote{Alternatively, the transfer might also be induced by spontaneous Rayleigh scattering from the side-mode atoms, similarly to the initial onset of superradiance where the two first-order side modes $(1,\pm 1)$ become seeded from the BEC. It would be of interest to compare the relative significance of the two alternatives.}. As discussed above, as soon as the side mode becomes slightly occupied, subsequent stimulated Rayleigh scattering into this side mode occurs under the production of resonant light since in this way energy conservation is ascertained, and a resonant Bragg scattering process ensues. In other words, superradiant Bragg scattering adapts itself to the occupation of higher-order side modes by adding resonant components to the frequency spectrum of the endfire mode radiation. The self-adaptation process is crucial, because otherwise higher-order side modes could never be occupied because of the increasing detuning barriers. 

\def\captionone{
Frequency spectra $|\tilde \cE_+(\omega)|^2$ of emitted endfire mode radiation. Calculations are for $g=3.5\times 10^5\,$s$^{-1}$ and pulse durations $210\,\mu$s (a), $350\,\mu$s (b), $1100\,\mu$s (c); $g=6.4\times 10^5\,$s$^{-1}$ and duration $750\,\mu$s (d). The insets show gray-scale representations of the side mode distributions (as explained in Fig.\ \ref{mwa:fig} and Sec.\ V) at the end of the pulse.
}
\LorP{}{
\begin{figure}[t]
\centerline{\includegraphics[width=8.0cm]{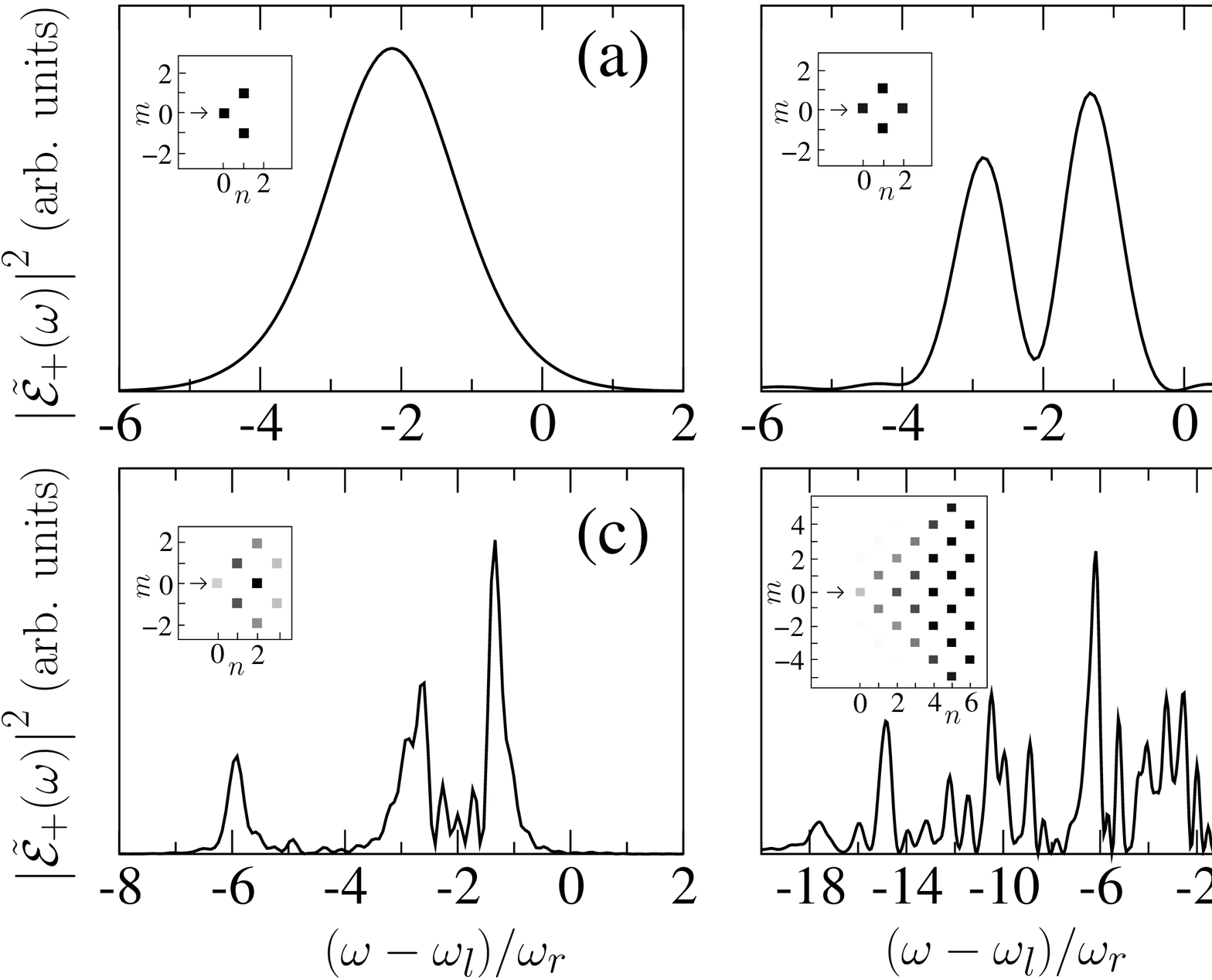}}
\caption{\captionone
\label{spectra}}
\end{figure}
}
\LorP{{\bf \dots\ Fig.\ 1 about here \dots}}{}

In Fig.\ \ref{spectra}, we illustrate this self-adaptation by calculating some spectra $|\tilde \cE_+(\omega)|^2$ for superradiant light emission with different applied laser pulses. Thereby, $\tilde \cE(\omega)$ denotes the temporal Fourier transform of $\cE_+(z,t)e^{-i(\omega t\mp kz)}$. The field envelope $\cE_+(z,t)$ is calculated with the help of Eqs.\ (\ref{sdeq}) for a rectangular applied laser pulse, and the Fourier transform is taken at some fixed $z_0$ outside the condensate. Figure \ref{spectra}(a) shows a situation where only the two first-order side modes have been occupied. We obtain a single broad peak centered at $\omega_l-2\omega_r$. As soon as the side mode $(2,0)$ becomes populated as well, the single peak splits [Fig.\ \ref{spectra}(b)]. This effect has already been analyzed in Refs.\ \cite{VasEfiTri04,TriSha04}. 
The situation of Fig.\ \ref{spectra}(c) corresponds to a case where the higher-order side modes $(2,\pm 2)$ and $(3,\pm 1)$  have been populated appreciably. We see that another frequency component at $\omega_l-6\omega_r$ arises. This shows that the endfire modes now also contain light resonant with the transition to these side modes. In case even higher-order modes become populated, two further pronounced peaks appear in the spectrum [Fig.\ \ref{spectra}(d)]. One is located at $\omega_l-10\omega_r$ as expected, whereas the other one is shifted slightly from the predicted value $\omega_l-14\omega_r$ to $\omega_l-15\omega_r$. However, we also find a stronger background, and the spectrum altogether becomes more complex.

\def\captiontwo{
Grouping of momentum side modes. The transition between two side modes is resonant with endfire-mode radiation of frequency $\omega_l+x\omega_r$ where $x$ is given by the number in the box. The resonance conditions imply detuning barriers that have to be overcome in the course of sequential superradiant scattering and lead to a separation of the side modes into several groups as indicated by the symbols.
}
\LorP{}{
\begin{figure}[t]
\centerline{\includegraphics[width=8.0cm,angle=270]{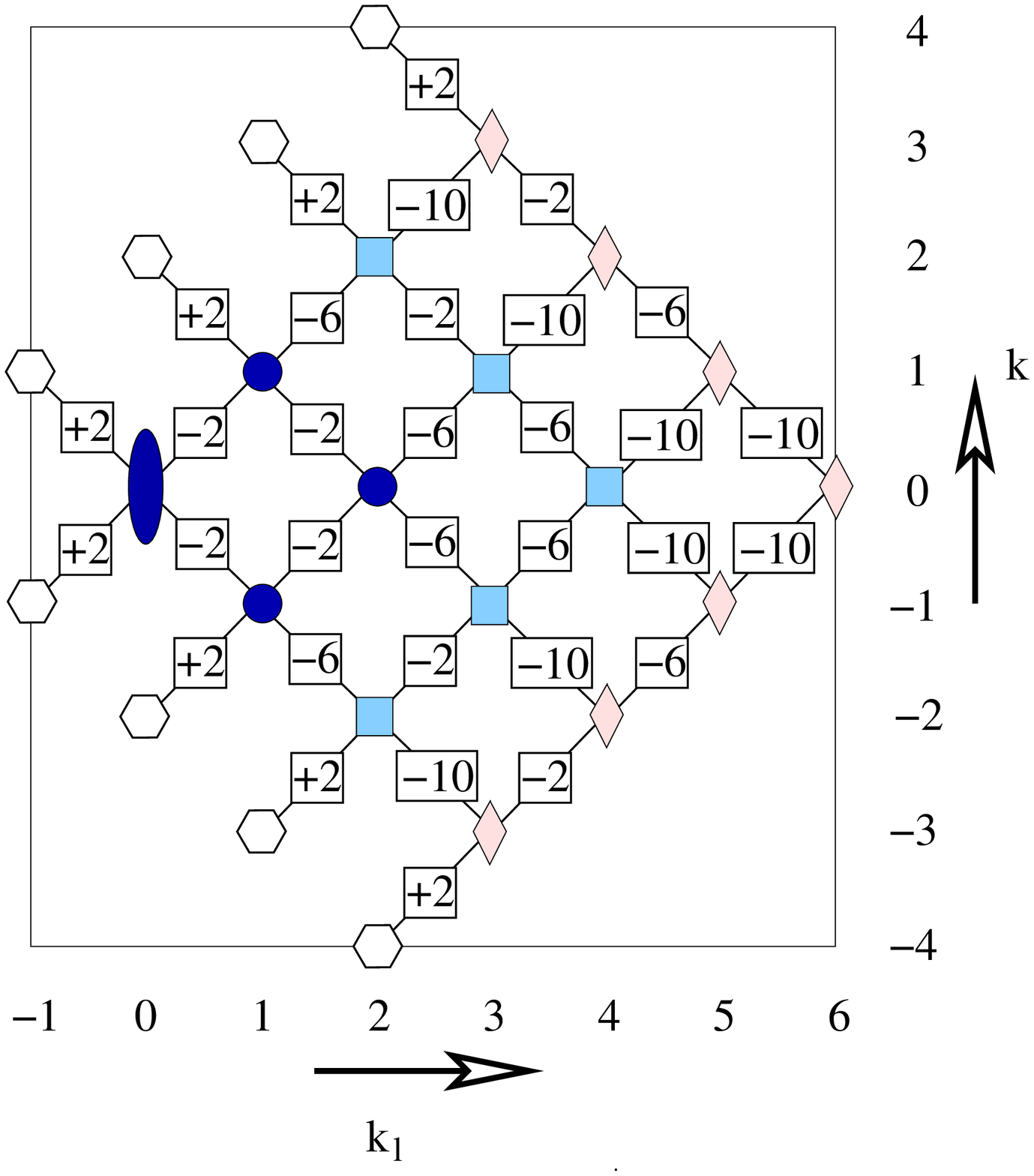}}
\caption{\captiontwo
\label{grid}}
\end{figure}
}
\LorP{{\bf \dots\ Fig.\ 2 about here \dots}}{}

{\it (ii) Detuning barriers.} Resonant Bragg scattering between the side modes $(n,m)$ and $(n+1,m\pm 1)$ requires endfire-mode radiation of frequency $\omega_l-2(n\pm m+1)\omega_r$. Figure \ref{grid} displays a selected set of side modes together with the resonance frequencies for transitions between adjacent modes. The figure shows that the resonance conditions imply the existence of several groups of side modes that are separated by detuning barriers. 
In the diagram, these groups are distinguished by different symbols. The first group consists of the BEC mode $(0,0)$ (large ellipse), the two first-order side modes $(1,\pm 1)$ and the mode $(2,0)$ (circles). Immediately after superradiant scattering has started up, the endfire modes only contain light of frequency $\omega_l-2\omega_r$. This light is resonant with the transitions within the first group. In other words, atoms can pass from the BEC to the side mode $(2,0)$ without having to overcome a detuning barrier. However, all possible transitions from the first group to the adjacent one that consists of the side modes $(2,\pm 2)$, $(3,\pm 1)$, and $(4,0)$ (circles) are resonant with light of frequency $\omega_l-6\omega_r$. This means that the two groups are separated by a detuning barrier. In our discussion of self-adapting Bragg scattering, we have explained how the initially empty side modes of the second group become populated and resonant radiation at frequency $\omega_l-6\omega_r$ is produced. However, this process requires some time so that the population of the second group sets in only after some delay.

The same reasoning applies to the build-up of population in the subsequent groups of side modes for which scattered light at frequency $\omega_l-10\omega_r$ (diamond symbols), $\omega_l-14\omega_r$ (not shown), etc.\ is required. The height of the detuning barrier between adjacent groups stays constant at $4\omega_r$.

{\it (iii) Directionality of population transfer.} The arguments given in (ii) do not explain why the side-mode group shown with hexagonal symbols does not become populated appreciably in typical experiments in the Bragg regime. All the side modes in this group would require resonant radiation of frequency $\omega_l+2\omega_r$ to become populated from adjacent side modes. This behavior is due to a directionality inherent in the population transfer between the side modes. This effect is most easily understood from the spatially-independent model of Eqs.\ (\ref{sieq}). There, one finds that the population $P_{nm}(t) = |C_{nm}(t)|^2$ of a side mode evolves according to
\bea\label{pop}
\frac{d P_{nm}(t)}{dt} &\sim& P_{n-1,m\pm 1}P_{nm} - P_{n+1,m\pm 1} P_{nm}\nonumber \\
&  &+{\rm\ additional\ terms}
\eea
To derive this relation, we have inserted Eqs.\ (\ref{e_p_si}) and (\ref{e_m_si}) into Eq.\ (\ref{env_psi_si}). 
The first two terms on the right-hand side of (\ref{pop}) arise from the interaction of the side mode $(n,m)$ with the resonant components $e^{i(n \pm m)\tau}C_{nm}C_{n+1,m\pm 1}^*$ and $e^{i(n\pm m-2)\tau}C_{n-1,m \mp 1}C_{nm}^*$ in the endfire mode radiation. They clearly indicate that population transfer from the side mode $(n,m)$ preferentially occurs to the modes $(n+1,m\pm 1)$ rather than to $(n-1,m\pm 1)$. Physically, this behavior is explained by the fact that a forward transfer to the side modes $(n+1,m\pm 1)$ amplifies the resonant endfire mode radiation and in this way the rate of subsequent Bragg scattering. Conversely, a backward transfer attenuates the endfire mode field. The additional terms in Eq.\ (\ref{pop}) that are not written out are either explicitly oscillating or at least do not possess a well-defined sign. They thus do not imply a preferred direction in the population transfer. Nevertheless, they play an important role in the overall process since, for example, they can initiate the population of empty side modes.

The concepts explained above can account for the appearance of the fan-shaped momentum patterns: The atomic side modes can be grouped into sets according to the detuning barriers implied by the resonant transition frequencies (Fig.\ \ref{grid}). In order to populate a new group a detuning barrier has to be overcome which only happens after some delay since resonant radiation has to be generated. The population of new groups thus occurs discontinuously in a stepwise process. However, due to the directionality of population transfer, only certain side-mode groups become occupied (filled symbols in Fig.\ \ref{grid}). The appearance of the fan patterns then becomes immediately obvious from the arrangement of the groups in Fig.\ \ref{grid}.

To illustrate the operation of the principles discussed above, it is very instructive to study the example presented in Figs.\ 13(a) and (b) of Ref.\ \cite{ZobNik06}, although the full fan pattern does not show up there. These simulations demonstrate the generic behavior of the side-mode dynamics for very weak pulses within the spatially independent model (see also \cite{Pio03}, in particular Fig.\ 5). Three groups of side modes are involved: the first one where all modes are temporarily populated, and the two subsequent ones where only the modes $(3,\pm 1)$, $(4,0)$ and $(5,\pm 1)$, $(6,0)$, respectively, become occupied. In particular, we see that population transfer {\it within} a group, e.g., from $(3,\pm 1)$ to $(4,0)$, occurs rather quickly once the modes $(3,\pm 1)$ become populated; however, the transport {\it between} groups, e.g., from $(4,0)$ to $(5,\pm 1)$ only happens after a delay during which resonant radiation is built up. For the same reason, population of, e.g., the modes $(4,\pm 2)$ that could be reached from $(3,\pm 1)$ does not occur. This transition is detuned so that the atoms rather undergo the resonant and fast transfer to $(4,0)$.
We also point out the temporal regularity in the transition process which also persists at longer times that are not displayed. For this regularity to appear it is necessary that the detuning barrier remains the same between different side mode groups. Furthermore, the example also clearly illustrates the directionality in the transitions since the atoms only undergo transition to higher-order side modes. In particular, the atoms rather ``wait" until they can overcome a detuning barrier instead of going backwards on a resonant transition. Finally, we note that full fan patterns are realized for somewhat stronger pump pulses. If we consider, e.g., the transition between the first and second side-mode group, radiation of frequency $\omega_l-6\omega_r$ can build up before all population has moved from the side modes $(1,\pm 1)$ to $(2,0)$. In this way, the modes $(2,\pm 2)$ become populated from the $(1,\pm 1)$ modes, and the full fan pattern starts to emerge.

We think that the mechanisms discussed in this section are essential for explaining and understanding sequential scattering in the Bragg regime. Nevertheless, the actual behavior found in the numerical simulations is rather complex, and the principles discussed here cannot account for all observable phenomena. For example, although we have shown in Ref.\ \cite{ZobNik06} that spatial effects are not essential for bringing about the fan-shaped distribution patterns, they may become significant in certain situations. In the case of very weak pulses that was discussed in the previous paragraph they lead to substantial deviations from the behavior found in the spatially independent model \cite{ZobNik06}. Furthermore, they also complicate the simple picture conveyed in Eq.\ (\ref{pop}) without, however, affecting the essential conclusion. The simulations also show that population can sometimes move backwards or oscillate between side modes.

\section{Transition between Kapitza-Dirac and Bragg regime}

The Kapitza-Dirac regime is characterized by the fact that individual atom-photon scattering events are not rigidly constrained by the condition of energy conservation. This means that the atoms may equally well scatter laser and endfire-mode photons leading to the simultaneous appearance of backward and forward recoiling atoms. In the Bragg regime, the atoms only possess forward-directed momenta, i.e., the predominant process is the scattering of laser radiation into the endfire modes in accordance with the rules of energy conservation. In Ref.\ \cite{RobPioBon04}, an interesting question was raised regarding the distinction between the two cases. It was suggested that the regime of scattering is determined by the strength of the atom-photon coupling $g$ rather than the duration of the pump laser as claimed in \cite{SchTorBoy03}.

\def\captionthree{
Ratio between backward and forward scattered atoms as a function of the pulse duration for $g=5\times 10^6\,$s$^{-1}$ (full curve), $3\times 10^6\,$s$^{-1}$ (dashed), $10^6\,$s$^{-1}$ (dotted), $3\times 10^5\,$s$^{-1}$ (dash-dotted). The inset shows this ratio for fixed field strength of the endfire modes $e_+=e_-= 4\times 10^{-3}\,$s$^{-1}$ (full curve), and $ 10^{-3}\,$s$^{-1}$ (dashed) with $g=3\times 10^6\,$s$^{-1}$.
}
\LorP{}{
\begin{figure}[t]
\centerline{\includegraphics[width=8.0cm]{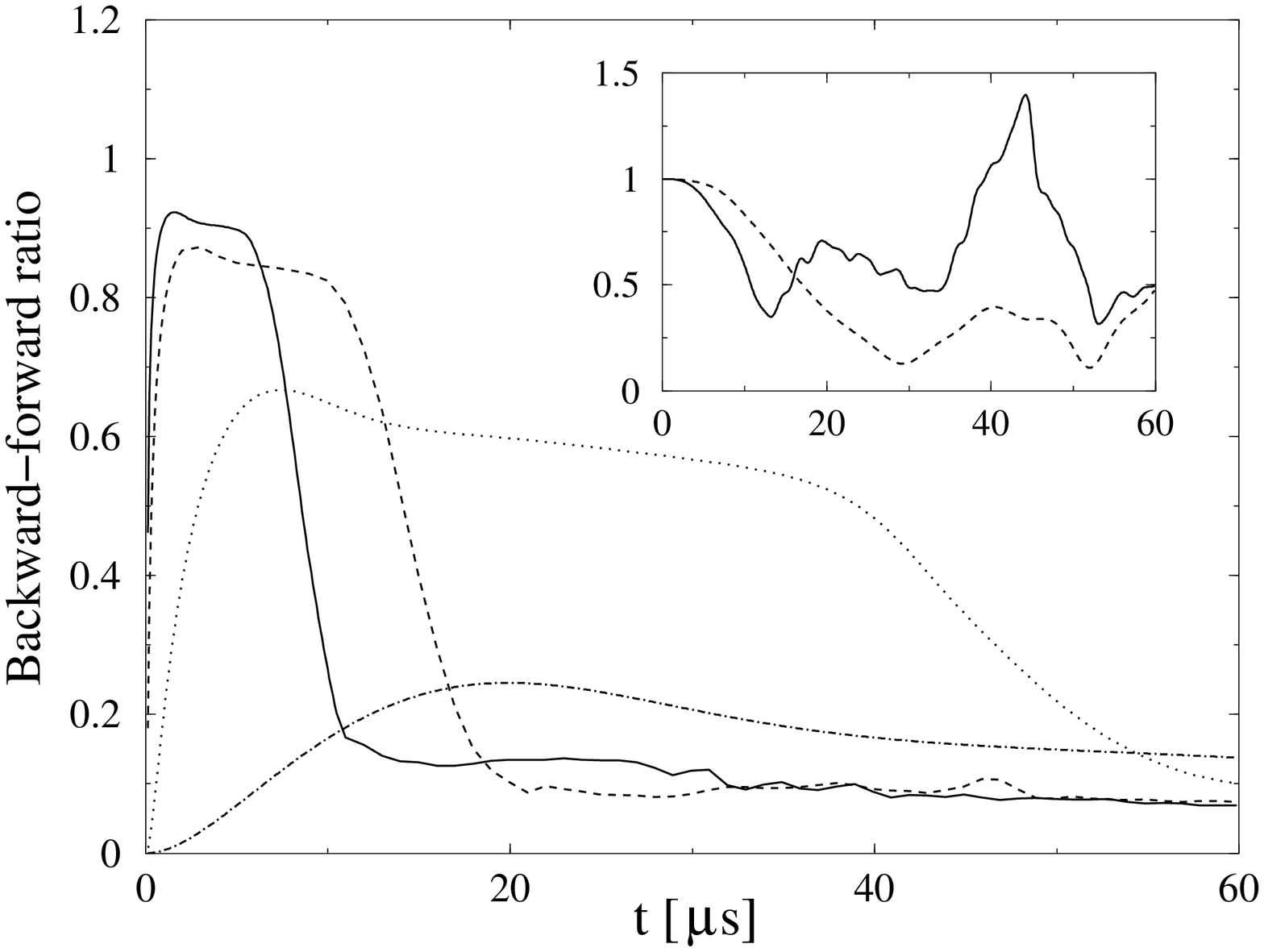}}
\caption{\captionthree
\label{strong}}
\end{figure}
}
\LorP{{\bf \dots\ Fig.\ 3 about here \dots}}{}

In this section, we would like to further explore this issue by means of numerical simulations with strong applied pulses of longer duration. Here, ``strong" refers to a large superradiant gain compared to the recoil frequency, whereas the scale for measuring the pulse duration is set by the recoil time. The main results of our studies are summarized in Fig.\ \ref{strong}. There we show the ratio between backward and forward scattered atoms, i.e., $\sum_{n<0,m} P_{n,m} / \sum_{n>0,m} P_{n,m}$ with $P_{n,m} = \int dz |\psi_{n,m}(z,t)|^2$, as a function of the pulse duration for different values of the coupling strength. By definition, this ratio should be close to 1 for the system in the Kapitza-Dirac regime. 
Based on these results we can draw the following conclusions within our model.
 
(i) The Kapitza-Dirac regime is only realized for strong {\it and} short pulses. With increasing pulse duration, the system enters the Bragg regime which can be seen from the sharp drop in the backward-to-forward ratio. 
We remark that in our simulations the X pattern builds up while the system is still clearly in the Kapitza-Dirac regime, in agreement with experimental results \cite{SchTorBoy03}. It is interesting to compare our observations regarding the transition between Kapitza-Dirac and Bragg regime to a case where the amplitudes $e_{\pm}(z,t)$ are kept constant. This corresponds to externally apply probe lasers along the BEC axis\footnote{For simplicity, we neglect probe depletion and photon scattering between the probe lasers. The intensities of the probe lasers are chosen such that the momentum spreads $\Delta p_{x}$ and $\Delta p_{z}$ are similar to the original simulations.}. In this case we observe a behavior displayed in the inset of Fig.\ \ref{strong}. We see that for sufficiently strong laser pulses the ratio between backward and forward scattered atoms remains much higher than in the original simulations and can even exceed one. This shows that the dynamic nature of the endfire modes plays a crucial role in the transition between the two regimes of scattering. 

(ii) The time that the system spends in the Kapitza-Dirac regime is {\it decreased} for growing laser intensity. This behavior might perhaps qualitatively be explained by the obervation that for growing coupling strength $g$, the system more quickly populates higher-order side modes that cover an increasing range $\Delta$ of detunings. The time scale $t\sim 1/\Delta$ over which nonresonant effects become noticeable is therefore shortened. 

(iii) For weak pulses the system is never in the Kapitza-Dirac regime, i.e., forward scattering always dominates over backward scattering, even at short times. This is in agreement with the analytical estimates of Ref.\ \cite{ZobNik06}.

We have confirmed the conclusions outlined above with a complementary analysis of the system dynamics. To this end, we have monitored the time evolution of ``effective rates" for photon emission and absorption to and from the optical endfire modes. This provides an alternative characterization of the two regimes of scattering, which of course is not completely independent of the one used above. The rates were calculated based the population transfer between the atomic side modes [the rates of population exchange between adjacent side modes can be obtained from Eq.\ (\ref{env_psi}), and, e.g., a net transfer from $(n,m)$ to $(n+1,m\pm 1)$ contributes to the total photon emission rate].
These studies show that in the strong-pulse regime, emission and absorption rates are almost equal at short times (Kapitza-Dirac regime). Simultaneous with the drop of the curves in Fig.\ \ref{strong}, however, emission starts to greatly exceed absorption. This clearly indicates the transition to the Bragg regime. Interestingly, with increasing pulse duration both emission and absorption rates go through a maximum, and subsequently the behavior becomes somewhat less clear-cut, although emission seems to always be stronger than absorption.

\def\captionfour{
Atomic side-mode distributions obtained within the spatially dependent 
model and unbalanced seeding. The gray level of each square represents the relative probability $\tilde p_{nm}$ as defined in the text. (a-b) Strong-pulse 
regime for $g=4.5\times 10^6\,{\rm s}^{-1}$ and pulse durations 
$t_{\rm p}=2.1\,\mu$s (a); 
$t_{\rm p}=3.4\,\mu$s (b).  (c-f) Weak-pulse 
regime for $g=4.5\times 10^5\,{\rm s}^{-1}$ and $t_{\rm p}=82.35\,\mu$s (c); 
$t_{\rm p}=103.9\,\mu$s (d); $t_{\rm p}=135.5\,\mu$s (e); 
$t_{\rm p}=192.6\,\mu$s (f). 
}
\LorP{}{
\begin{figure}[t]
\centerline{\includegraphics[width=8.0cm]{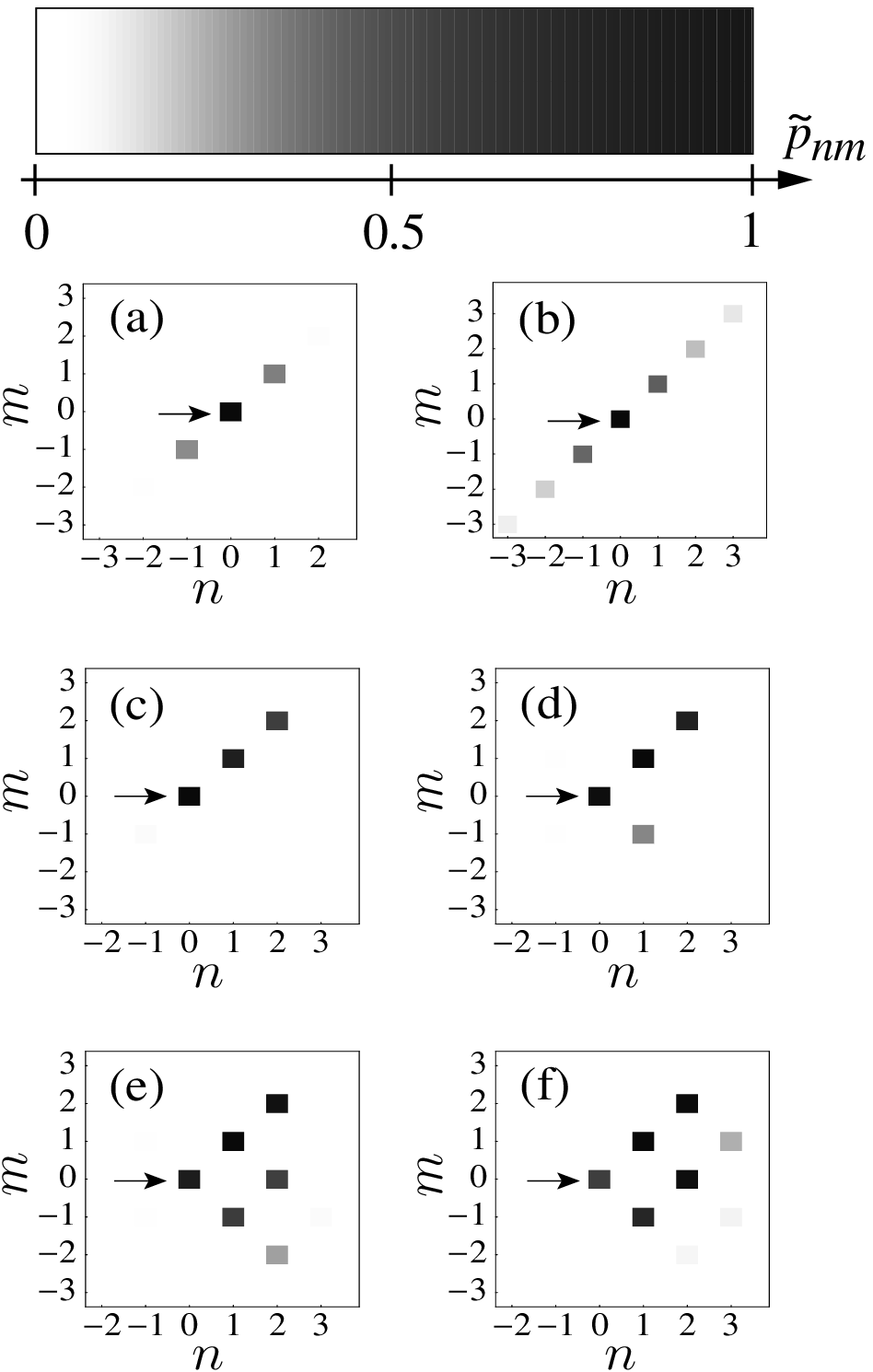}}
\caption{\captionfour
\label{mwa:fig}}
\end{figure}
}

\section{Sequential scattering with seeded side modes}
So far we have concentrated on superradiant scattering which is initiated 
by quantum mechanical noise, i.e., spontaneous Rayleigh scattering which occurs in 
the absence of any side-mode populations. 
Experimentally, this spontaneous startup of the phenomenon can be strongly suppressed 
by seeding, for instance, one of the first-order atomic side modes. 
Such a preparation can be achieved by initially exposing the BEC 
to a pulsed optical standing wave which transfers a small fraction of 
the atoms from the BEC to the desired side-mode 
[let us say $(1,1)$] by Bragg diffraction \cite{Koz99}. 
In this case, subsequent application of a pump pulse leads to a very rapid 
growth of the initially occupied side mode due to bosonic enhancement 
whereas the population transfer to the other first-order side 
mode $(1,-1)$, is much slower since it is still initiated by 
quantum mechanical noise. As a result, at the end of the pump pulse 
the main part of the population has been transferred 
from the BEC $(0,0)$ to the side mode $(1,1)$, provided that higher-order 
side modes are not populated \cite{InoPfaGup99,SchCamStr04,KozSuzTor99}. 
We see therefore that in the case of seeding, superradiant scattering 
essentially leads to amplification of the initially prepared weak 
matter-wave side mode. Most importantly, it has been shown that this matter-wave 
amplification is phase coherent \cite{InoPfaGup99,SchCamStr04,KozSuzTor99}.

\LorP{{\bf \dots\ Fig.\ 4 about here \dots}}{} 
 
In this section, we would like to discuss superradiant Rayleigh scattering 
initiated by seeding one of the first-order side modes. In particular, 
our purpose is to explore its effect on the distribution of the atomic side modes in sequential superradiant scattering. 
To this end we have solved the spatially-dependent Eqs.\ (4), with various 
unbalanced seed functions $\psi_{1,-1}(z,0)\neq\psi_{1,1}(z,0)$. 
In Figs. \ref{mwa:fig}, we present the resulting side-mode distribution patterns for both strong and 
weak-pulse regimes with seed functions 
$\psi_{1,-1}(z,0)=\psi_{0,0}(z,0)/\sqrt{N}$ and 
$\psi_{1,1}(z,0)=\sqrt{500}\psi_{0,0}(z,0)/\sqrt{N}$. The gray level of each square represents the ``relative" probability $\tilde p_{nm}=P_{nm}/P_{nm}^{\rm (max)}$ with $P_{nm} = \int dz |\psi_{nm}(z,t)|^2$ and $P_{nm}^{\rm (max)} = {\rm max}_{\{n,m\}}P_{nm}$.

In general, the stimulated nature of the startup is evidenced in two 
respects.  First, the time scales are much shorter than 
in the case of spontaneous startup 
(e.g., compare to Figs.\ 3 and 11 of Ref. \cite{ZobNik06}).  
Second, the side modes coupled to the initially populated 
$(1,1)$ side mode become populated much faster than other side modes.
Hence, as depicted in Figs.\ \ref{mwa:fig}(a,b), in the strong-pulse 
regime only 
diagonal side modes $(n,n)$ become populated. Such a distribution might  
be useful for the construction of a bidirectional correlated atom laser. 
More precisely, it is obvious that if the strong pump pulse is 
followed by a free evolution for a time $t_{\rm f}$, 
one obtains a sequence of atomic clouds (pulses) propagating in opposite 
directions and separated by distances $\sim \sqrt 2\hbar k_l t_{\rm f}/M$. 
A ``seeded" realization of the atom laser might have several advantages compared to the spontaneous start-up. First, one has a much greater control over the time evolution of the process since it is not initiated by fluctuations. Second, the effect of potentially detrimental collisions between side modes $(n,n)$ and $(n,-n)$ emanating in the forward direction (see Fig.\ 2(a) of \cite{ZobNik06}) is significantly reduced due to the imbalance in the populations.

In the weak-pulse regime the atomic side-mode distributions are more 
involved, in general [see Figs. \ref{mwa:fig}(c)-(f)]. For relatively short times only side 
modes in the forward upper diagonal $(n,n)$ are populated. This is because the field of the ${\cal E}_+$ endfire mode initially is too weak to induce appreciable population of the $(2,0)$ or other off-diagonal side modes (compare with discussion in Sec.\ V.A of \cite{ZobNik06}). 
Similar to the strong-pulse regime [Figs. \ref{mwa:fig}(c)-(f)], we thus typically have new types of patterns in the sense that they are clearly asymmetric 
with respect to the laser direction contrary to the case of spontaneous 
startup where they were almost symmetric. The unbalanced seeding has broken 
this symmetry. However, for sufficiently large times bosonic stimulation 
starts working efficiently also for the spontaneously initiated side 
mode $(1,-1)$ and the side modes coupled to it become populated.  
As a result we observe a redistribution of the population and the 
appearance of more symmetric patterns resembling the well known fan-shape 
observed in the case of spontaneous startup 
(compare to Figs. 11 of Ref. \cite{ZobNik06}). The re-emergence of more symmetric patterns appears to be more pronounced than in the strong-pulse regime due to the longer time scales involved.

\section{Summary}
In this paper we have discussed several aspects of sequential superradiant scattering from atomic Bose-Einstein condensates. Our treatment was based on a semiclassical description of the process in terms of the coupled Maxwell-Schr\"odinger equations for the matter-wave and optical fields, developed in Refs.\ \cite{ZobNik05,ZobNik06}. First of all, we investigated sequential scattering in the Bragg regime. We identified three mechanisms which we think essential for understanding the system behavior in this regime. A first element are the effective detuning barriers between the atomic side modes. Although at first sight the arrangement of these barriers already seems to imply the formation of fan-shaped side-mode patterns (see Fig.\ \ref{grid}), the actual appearance of these patterns can only be explained conclusively by taking two further mechanisms into account: on the one hand, ``self-adaptive" Bragg scattering by which the system can adapt the frequency spectrum of the scattered endfire mode radiation such that the transitions between atomic side modes remain resonant. If it weren't for this effect, the atoms would not be able to occupy higher-order side modes, in particular for weaker pump pulses. On the other hand, the population transfer between side modes preferentially takes place along the direction of the pump laser, rather than against it. This aspect makes sure that only certain atomic side modes can become populated in the Bragg regime.
We also pointed out that the system dynamics may become quite complex at times and further factors, e.g., spatial effects, may have to be taken into consideration to fully account for the observed behavior. However, rather than trying to theoretically analyze all these effects in detail, it would be more desirable to have further experimental data available to refine the models and to gain more insight into the system. For example, if a stepwise population of the different side-mode groups as described in Sec.\ III could be verified experimentally, this would somewhat favor an interpretation of the superradiant process in terms of atomic diffraction from optical gratings, rather than photon diffraction from matter-wave gratings (cf.\ discussion in \cite{SchTorBoy03}).

As a second topic, we examined some aspects of the relationship between the Kapitza-Dirac and Bragg regimes of superradiant scattering. Within our model, we find that for weak applied pulses, the system is always in the Bragg regime, even at times short compared to the recoil time. This is of course already expected from analytical models of the short-time dynamics \cite{ZobNik06,RobPioBon04}. In other cases, however, analytical descriptions are not feasible, and numerical methods have to be applied. In particular, when investigating stronger pulses, we find that the system always makes a transition from the Kapitza-Dirac to the Bragg regime for increasing pulse duration. Interestingly, the time spent in the Kapitza-Dirac regime seems to decrease for growing pulse duration.

Finally, we also investigated how the initial seeding of one of the atomic side modes by way of Bragg diffraction influences the sequential scattering process. Questions of this kind can be expected to be of interest in connection with studies of matter-wave amplification \cite{InoPfaGup99,SchCamStr04,KozSuzTor99}. We presented typical examples for atomic side mode distributions in the weak- and strong-pulse regimes. In particular, we find new patterns in the sense that they are clearly asymmetric 
with respect to the laser direction contrary to the case of spontaneous 
startup. This asymmetry is of course reflective of the preparation of the initial state. The extent of asymmetry, however, can vary appreciably in the course of time, especially in the weak-pulse regime.

GMN acknowledges support by ``Pythagoras II'' of the EPEAEK 
research programme.

\LorP{
\begin{widetext}
\newpage
\begin{figure}[t]
\centerline{\includegraphics[width=25.0cm,angle=270]{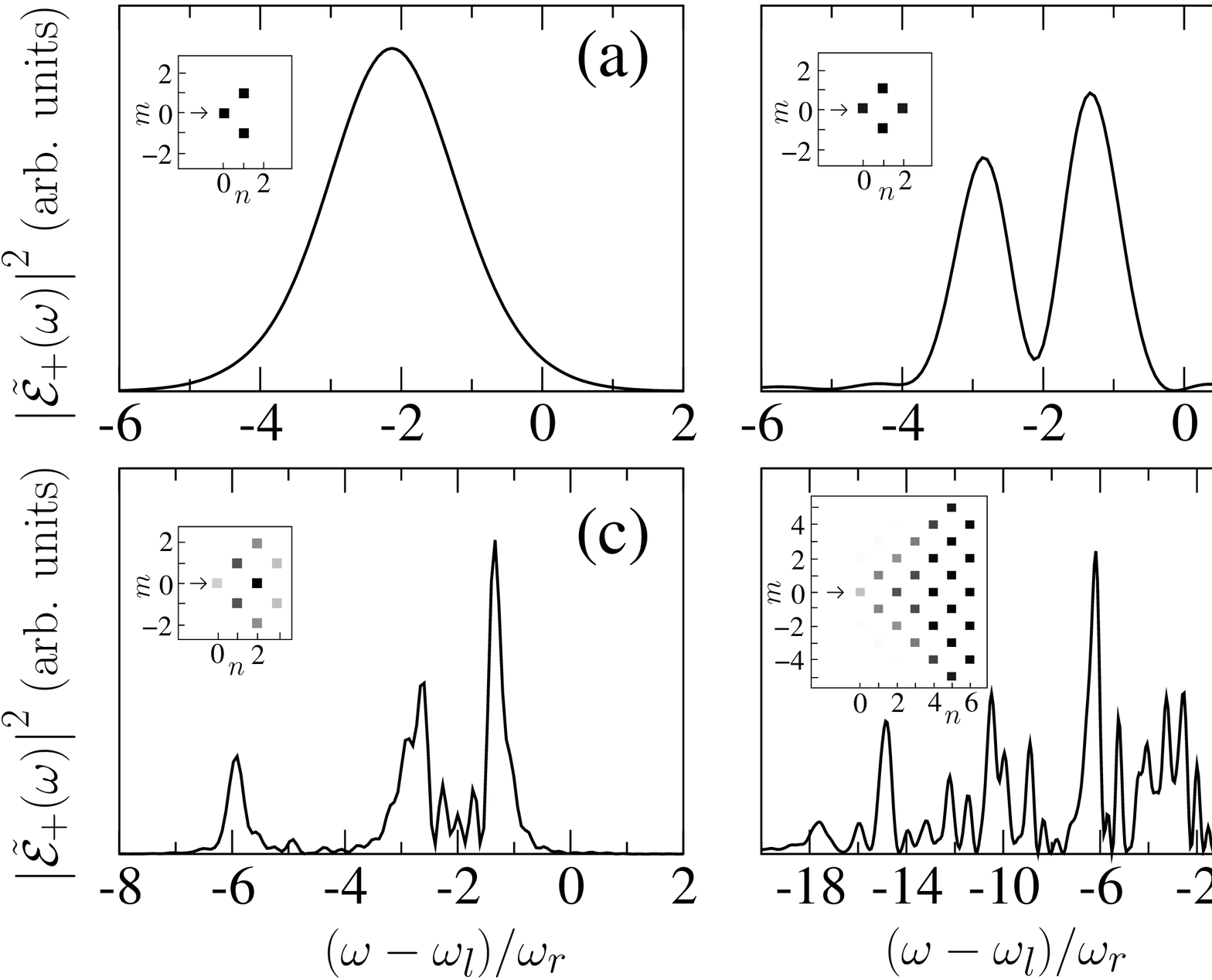}}
\caption{\captionone
\\[0.2cm] {\bf Paper: Zobay, Sequential superradiant scattering from atomic Bose-Einstein condensates}
\label{spectra}}
\end{figure}
\newpage
\begin{figure}[t]
\centerline{\includegraphics[width=21.0cm,angle=270]{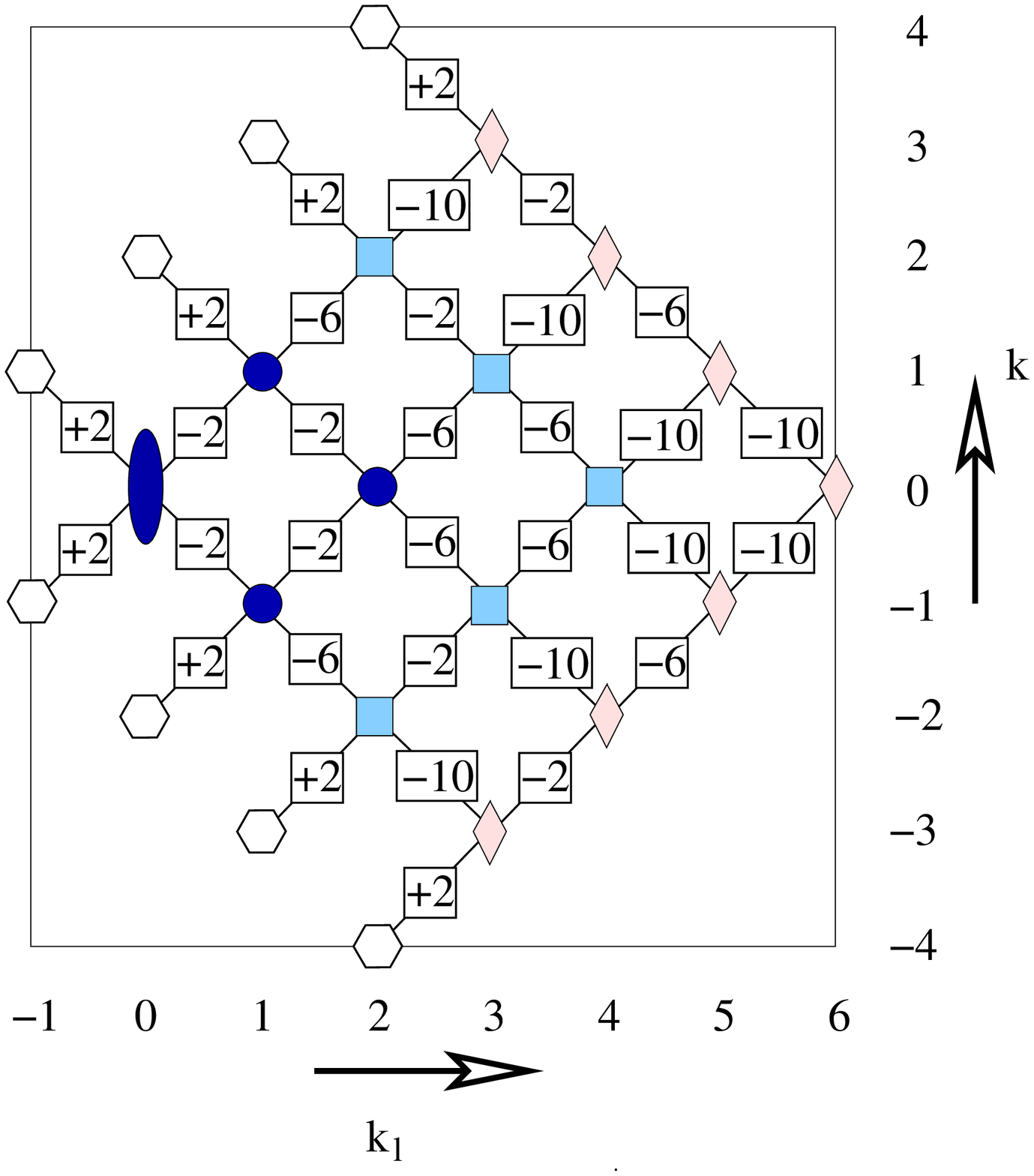}}
\caption{\captiontwo
\\[0.2cm] {\bf Paper: Zobay, Sequential superradiant scattering from atomic Bose-Einstein condensates}
\label{grid}}
\end{figure}
\newpage
\begin{figure}[t]
\centerline{\includegraphics[width=23.0cm,angle=270]{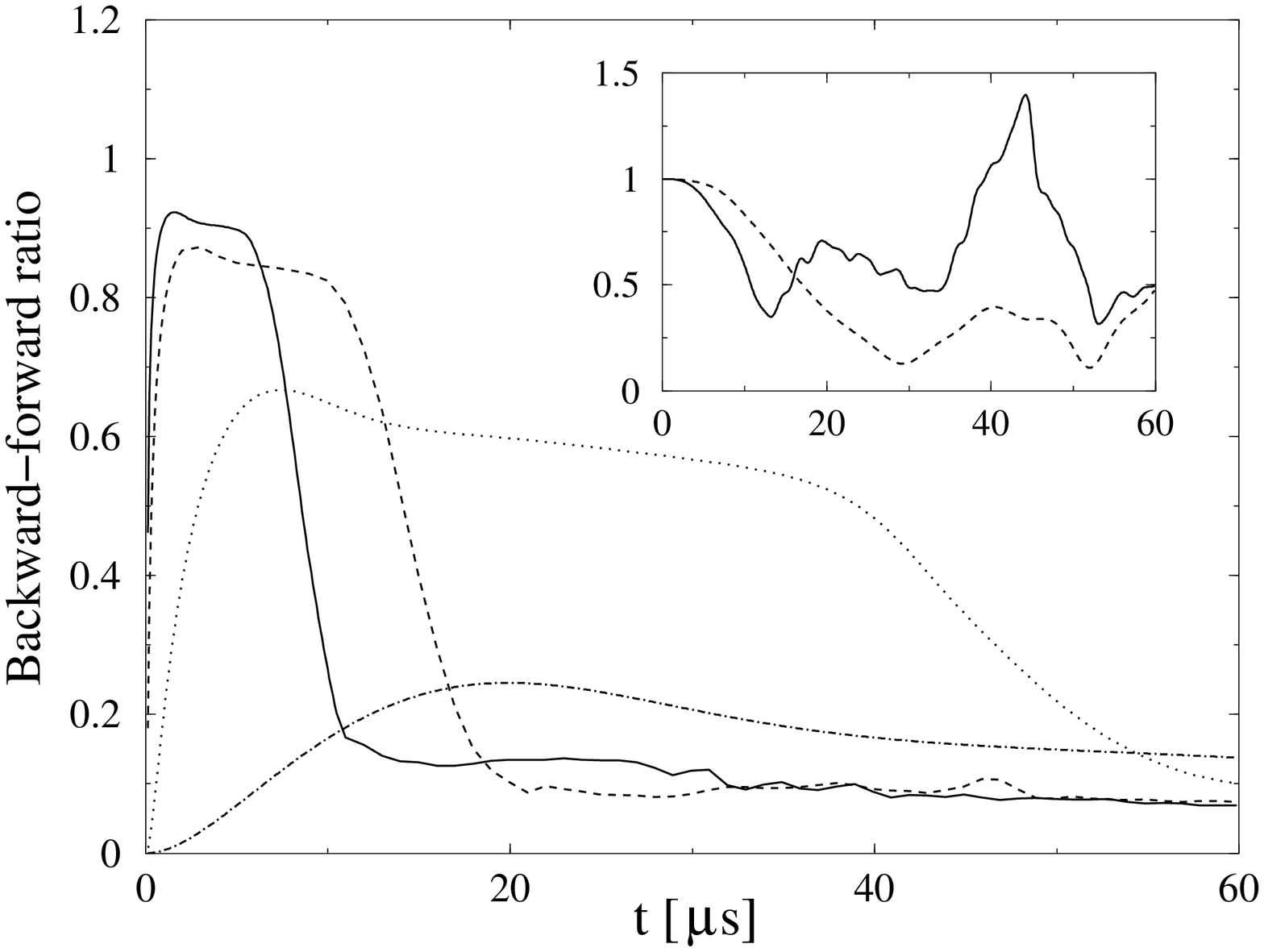}}
\caption{\captionthree
\\[0.2cm] {\bf Paper: Zobay, Sequential superradiant scattering from atomic Bose-Einstein condensates}
\label{strong}}
\end{figure}
\begin{figure}[t]
\centerline{\includegraphics[width=15.0cm]{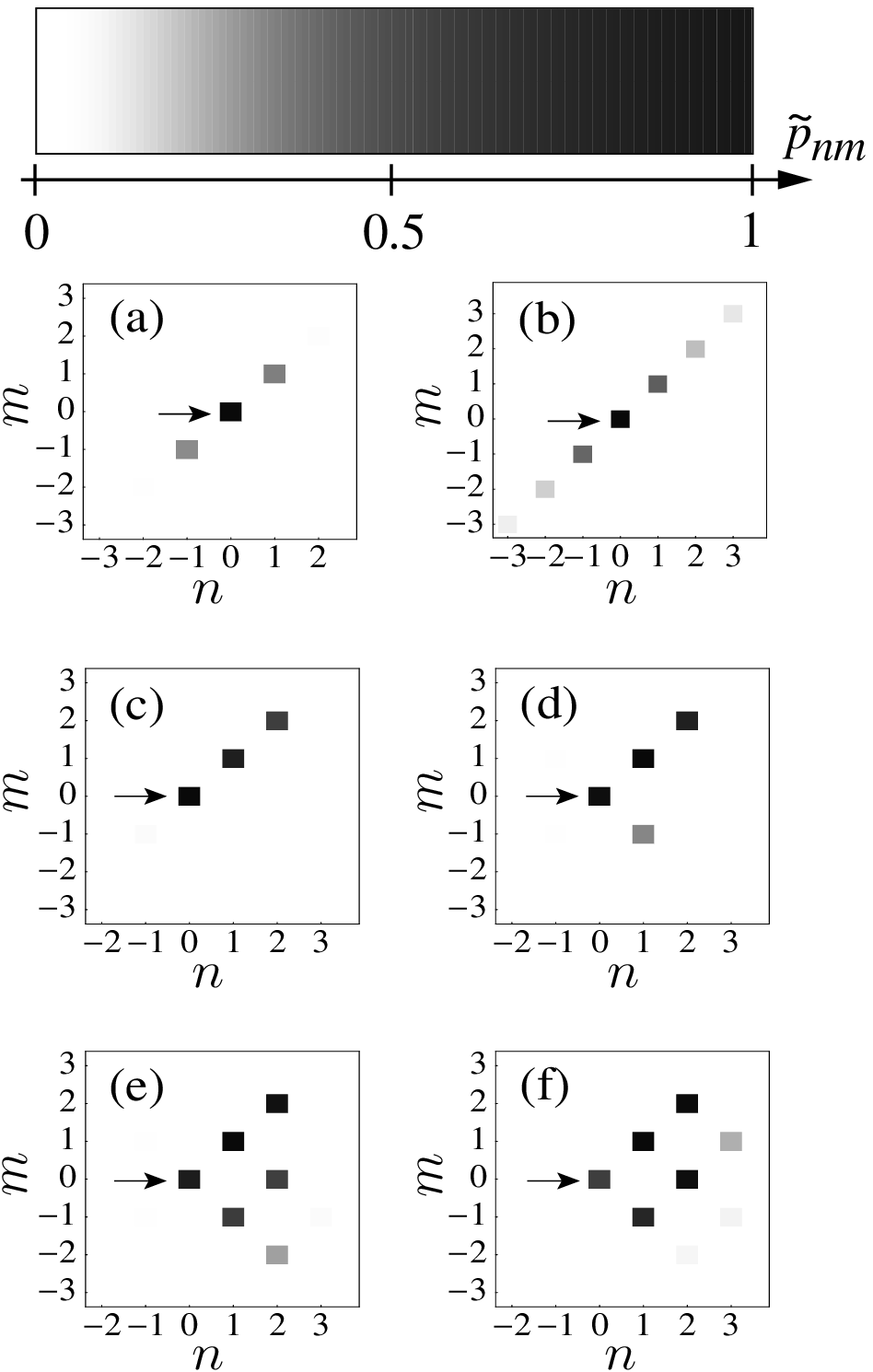}}
\caption{\captionfour
\\[0.2cm] {\bf Paper: Zobay, Sequential superradiant scattering from atomic Bose-Einstein condensates}
\label{mwa:fig}}
\end{figure}
\end{widetext}
\newpage
\begin{figure}[t]
{\bf
\noindent \underline{Corresponding author information:}\\[0.3cm]
O. Zobay \\[0.1cm]
Institut f\"ur Angewandte Physik\\ 
Technische Universit\"at Darmstadt\\ 
Hochschulstr.\ 4a\\
64289 Darmstadt\\
Germany\\[0.1cm]
Tel.\ (office): +49 6151 163179\\
Tel.\ (home): +49 6151 4291120\\
Fax: +49 6151 163279\\[0.1cm]
E-Mail: oliver.zobay@physik.tu-darmstadt.de
}
\end{figure}
}{}
\end{document}